\documentclass[fleqn,twoside]{article}
\usepackage{espcrc2}

\usepackage{graphicx}
\usepackage[figuresright]{rotating}

\newcommand{\AmS}{{\protect\the\textfont2
  A\kern-.1667em\lower.5ex\hbox{M}\kern-.125emS}}

\newcommand{\beqn}{\begin{eqnarray}}
\newcommand{\eeqn}{\end{eqnarray}}
\newcommand{\beqns}{\begin{eqnarray*}}
\newcommand{\eeqns}{\end{eqnarray*}}

\newcommand{\intl}{\int\limits}

\newcommand{\qqbar}{q\overline{q}}

\def\ea{{\it et al.}}
\def\GeVM{~${\rm GeV}/c^2$}

\def\GeVp{~${\rm GeV}/c$}
\def\GeVE{~${\rm GeV}$}

\title{Measurement of the $e^+e^-\to$ hadrons cross-section at low energy with ISR events at BABAR}

\author{B. MALAESCU\address[MCSD]{Laboratoire de l'Acc{\'e}l{\'e}rateur Lin{\'e}aire, \\ 
        IN2P3/CNRS, Universit\'e Paris-Sud 11, Orsay, France}
	\thanks{Now at CERN, CH--1211, Geneva 23, Switzerland.} \\ 
        On behalf of the BABAR Collaboration
        } 
       
\begin{document}

\begin{abstract}
The precise measurement of the cross section $e^+e^-\to\pi^+\pi^-(\gamma)$ from threshold to an energy of $3$\GeVE, 
using events with Initial State Radiation~(ISR) collected with the BABAR detector, is presented. 
The ISR luminosity is determined from a study of the leptonic process $e^+e^-\to\mu^+\mu^-\gamma(\gamma)$, and the
method is tested by the comparison with the next-to-leading order (NLO) QED prediction. 
The leading-order hadronic contribution to the muon magnetic anomaly calculated using the BABAR $\pi\pi$ 
cross section measured from threshold to $1.8$\GeVE~ is
$(514.1 \pm 2.2({\rm stat}) \pm 3.1({\rm syst}))\times 10^{-10}$. 
Other results on ISR multihadronic cross sections from BABAR are presented. 
\vspace{1pc}
\end{abstract}

\maketitle

\section{Introduction}

The measurements of the $e^+e^-\to$ hadrons cross-section are used to evaluate dispersion integrals for 
calculations of the hadronic vacuum polarization~(VP). 
In particular, the hadronic contribution to the muon magnetic moment anomaly~($a_\mu^{had}$) requires data in 
the low mass region, dominated by the process $e^+e^-\to\pi^+\pi^-(\gamma)$ which provides $73\%$ of the contribution.
The dominant uncertainty comes also from the $\pi\pi$ channel. 

The systematic precision of the (recent) previously measured $\pi\pi$ cross sections is of $0.8\%$ for CMD2~\cite{cmd-2} and 
$1.5\%$ for SND~\cite{snd}, the two measurements being in good agreement. 
The first ISR measurement done by KLOE~\cite{kloe04} had a quoted systematic precision of $1.3\%$, and some deviation in 
shape was observed when comparing to the Novosibirsk data. 
For the reanalysed KLOE data~\cite{kloe08} a systematic uncertainty of $0.9\%$ is quoted, and the agreement with the 
Novosibirsk data is improved. 

The comparison of the theoretical and measured~\cite{bnl} values of $a_\mu$ shows a discrepancy of about $3\sigma$ 
when previous $e^+e^-$ data~\cite{cmd-2,snd,kloe08} are used, possibly hinting at new physics. 
An approach using $\tau$ decay data corrected for isospin-breaking, leads to a smaller difference~\cite{newtau}. 

The BABAR $2\pi$ results presented in these proceedings were published in \cite{:2009fg}. 
Their achieved goal was to obtain a measurement of the contribution of the $2\pi$ channel to ($a_\mu^{had}$) with a precision 
better than $1\%$, implying a control of systematic uncertainties at the $10^{-3}$ level.

\section{The BABAR ISR $\pi\pi$ analysis} 

The results on the $\pi\pi$ cross section presented here are obtained with the ISR method~\cite{isr} using $e^+e^-$ annihilation 
events collected with the BABAR detector, at a center-of-mass energy $\sqrt{s}$ near $10.58$\GeVE. 
We consider events $e^+e^- \to X\gamma_{ISR}$, where $X$ includes any final state, and the ISR~photon is emitted by the $e^+$ or $e^-$. 
The $e^+e^- \to \pi\pi(\gamma_{FSR})$ cross section is obtained as a function of $\sqrt{s'}$, which is the invariant mass of the final 
state. 
The advantage of the ISR method is that all the mass spectrum is covered at once~(from threshold to $3$\GeVE~ in BABAR)
with the same detector conditions and analysis. 

In the BABAR analysis the $\pi^+\pi^-\gamma_{ISR}(\gamma_{FSR})$ and $\mu^+\mu^-\gamma_{ISR}(\gamma_{FSR})$ spectra are measured. 
This is the first NLO measurement, an eventual additional radiation being taken into account in the analysis, instead of being corrected 
a posteriori (as done by previous experiments). 
The muon spectrum is compared with the NLO QED prediction, this important cross check of the analysis being called the QED test. 
The $\sqrt{s'}$ spectrum of $e^+e^-\to X\gamma$ events is related to the cross section for the process $e^+e^-\to X$ through
\begin{equation}
\label{def-lumi}
  \frac {\mathrm{d}N_{X\gamma}}{\mathrm{d}\sqrt{s'}}~=~\frac {\mathrm{d}L_{ISR}^{eff}}{\mathrm{d}\sqrt{s'}}~
    \varepsilon_{X\gamma}(\sqrt{s'})~\sigma_{X}^0(\sqrt{s'})~,
\end{equation}
where $\varepsilon_{X\gamma}$ is the detection efficiency (acceptance) determined by simulation with corrections obtained from data, 
and $\sigma_X^0$ is the bare cross section (excluding VP). 
The muon spectrum is used to derive the effective ISR luminosity $\rm{L}^{\rm eff}_{\rm ISR}$. 
We correct for the leading order FSR contribution for muons (smaller than $1\%$, below $1$\GeVE), while additional FSR photons are measured. 
The $\pi\pi(\gamma_{FSR})$ cross section is obtained from the ratio of the $2\pi$ spectrum and $\rm{L}^{\rm eff}_{\rm ISR}$, 
where the $e^+e^-$ luminosity, additional ISR effects, vacuum polarization and ISR photon efficiency cancel, 
hence the strong reduction of the systematic uncertainty. 

This analysis is based on $232~{\rm fb}^{-1}$ of data recorded with the BABAR 
detector~\cite{detector} at the PEP-II asymmetric-energy $e^+e^-$ storage 
rings. Charged-particle tracks are measured with a five-layer double-sided 
silicon vertex tracker (SVT) together with a 40-layer drift chamber (DCH) 
inside a 1.5 T superconducting solenoid magnet. The energy and direction
of photons are measured in the CsI(Tl) electromagnetic calorimeter (EMC).
Charged-particle identification (PID) uses ionization loss ${\rm d}E/{\rm d}x$
in the SVT and DCH, the Cherenkov radiation detected in a ring-imaging
device (DIRC), and the shower deposit in the EMC ($E_{cal}$) and in the 
instrumented flux return (IFR) of the magnet.

Two-body ISR events are selected by requiring a photon with $E_\gamma^*>3$\GeVE~ and 
laboratory polar angle in the range $0.35-2.4~{\rm rad}$, 
and exactly two tracks of opposite charge, each with 
momentum $p>1$\GeVp~ and within the angular range $0.40-2.45~{\rm rad}$. 
Events with one single charged track are also recorded and exploited for efficiency measurements. 

Signal and background ISR processes are simulated with Monte Carlo (MC) event
generators based on Ref.~\cite{eva}. Additional ISR photons are generated with
the structure function method~\cite{struct-fct}, and additional FSR photons 
with {\small PHOTOS}~\cite{photos}. 
The response of the BABAR detector is simulated with {\small GEANT4}~\cite{geant}. 

Background events from $e^+e^-\to\qqbar$ ($q=u,d,s,c$) are generated with {\small JETSET}~\cite{jetset}. 
They are due to low-multiplicity events in which an energetic $\gamma$ originating from a $\pi^0$ is mistaken as the ISR photon candidate. 
To normalize this rate from {\small JETSET}, the $\pi^0$ yield (obtained by
pairing the ISR photon with other photons in the event) is compared in data and MC. 
Multi-hadronic ISR backgrounds are dominated by
$e^+e^-\to\pi^+\pi^-\pi^0\gamma$ and $e^+e^-\to\pi^+\pi^-2\pi^0\gamma$ 
contributions. An approach similar to that for $\qqbar$ is followed to 
calibrate the background level from the $3\pi$ ISR process,
using $\omega$ and $\phi$ signals.
The MC estimate for the $2\pi 2\pi^0\gamma$ process is
used and assigned a 10\% systematic uncertainty.
Background contributions to the $\mu\mu$ spectrum are found to be negligible.

Acceptance and mass-dependent efficiencies for trigger, reconstruction, 
PID, and event selection are computed using the simulation.
The ratios of data and MC efficiencies have been determined from specific 
studies, as described below, and are applied as mass-dependent 
corrections to the MC efficiency. They amount to at most a few 
percent and are known to a few permil level or better.

Tracking and PID efficiencies are determined 
taking advantage of pair production. For tracking studies, two-prong ISR 
candidates are selected on the basis of the ISR photon and one track.  
A kinematic fit yields the expected parameters of the second track. 
The unbiased sample of candidate second tracks is used to measure track reconstruction efficiency.
The study of 2-particle overlap in the detector required a large effort to reach 
the per mil accuracies.

Each event is subjected to two kinematic fits to the $e^+e^-\to X\gamma$ hypothesis,
where $X$ allows for possible additional radiation.
Both fits use the ISR photon direction and the parameters and
covariance matrix of each charged-particle track.
The two-constraint (2C) `ISR' fit allows an undetected photon 
collinear with the collision axis, while the 3C `FSR' fit is performed only when an additional photon is detected. 
Most events have small $\chi^2$ values for both fits; an event with only a
small $\chi^2_{ISR}$ ($\chi^2_{FSR}$) indicates the presence of additional 
ISR (FSR) radiation. Events where both fits have large $\chi^2$ values 
result from track or ISR photon resolution effects, the presence of 
additional radiated photons, or multi-hadronic background.
To accommodate the expected background levels, different criteria in the 
($\chi^2_{ISR}$,$\chi^2_{FSR}$) plane are applied, i.e.  a loose 2D cut in the central $\rho$ region and a tighter cut for the $\rho$ tails. 
The loose cut is also used in the $\mu\mu\gamma$ analysis. 
The $\pi\pi$ ($\mu\mu$) mass is calculated from the best `ISR' or `FSR' fit.

The computed acceptance and the $\chi^2$ selection efficiency depend on the description of radiative effects in the generator. 
The difference of the FSR rate between data and MC is measured, resulting in a small correction for the cross section. 
More significant differences are found between data and the generator for additional ISR photons, since the latter
uses a collinear approximation and an energy cut-off for very hard photons. 
Induced kinematical effects have been studied using the NLO {\small PHOKHARA} generator~\cite{phok} at four-vector
level with fast simulation. 
Differences occuring in acceptance yield corrections to the QED test. 
In contrast, since radiation from the initial state is common to the pion and muon channels, 
the $\pi\pi(\gamma)$ cross section, obtained from the
$\pi\pi$/$\mu\mu$ ratio, is affected and corrected only at a few permil level.
The $\chi^2$ selection efficiency determined from muon data 
applies to pions, after correction for the effect of secondary
interactions and the $\pi/\mu$ difference for additional FSR. 
The measured $\pi\pi(\gamma)$ cross section is to a large extent insensitive to the description of NLO 
effects in the generator.

\section{The QED test} 

\begin{figure}[tb]
  \centering
  \includegraphics[width=8.0cm]{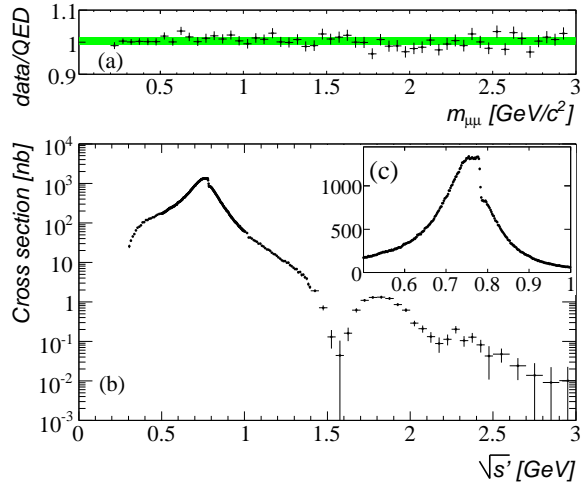}
  \caption{\small
(a) The ratio of the measured cross section for 
$e^+e^-\to\mu^+\mu^-\gamma(\gamma)$ to the NLO QED prediction. 
The band represents Eq.~(\ref{qed-test}).
(b) The measured cross section for $e^+e^-\to\pi^+\pi^-(\gamma)$ 
from 0.3 to $3$\GeVE. 
(c) Enlarged view of the $\rho$ region in energy intervals of 2 MeV.
The plotted errors are from the sum of the diagonal elements of the 
statistical and systematic covariance matrices.}
  \label{babar-log}
\end{figure}

The QED test involves two additional factors, both of which 
cancel in the $\pi\pi$/$\mu\mu$ ratio: $L_{ee}$ and the ISR photon efficiency, 
which is measured using a $\mu\mu\gamma$ sample selected only on the basis 
of the two muon tracks. The QED test is expressed as the ratio  
of data to the simulated spectrum, after the latter is corrected using data 
for all known detector and reconstruction differences. The generator is 
also corrected for its known NLO deficiencies using the comparison to 
{\small PHOKHARA}. The ratio is consistent with unity from threshold to 
$3$\GeVM, (Fig.~\ref{babar-log} (a)). A fit to a constant value yields 
($\chi^2/n_{\rm{df}}=55.4/54$; $n_{\rm{df}}$=number of degrees of freedom)
\begin{equation}
\label{qed-test}
 \frac {\sigma_{\mu\mu\gamma(\gamma)}^{data}} {\sigma_{\mu\mu\gamma(\gamma)}^{NLO~QED}}~-~1~=~ (40\pm20\pm55\pm94)\times 10^{-4}~,
\end{equation}
where the errors are statistical, systematic from this analysis, and 
systematic from $L_{ee}$ (measured using Bhabha scattering events), 
respectively. The QED test is thus satisfied within an overall accuracy of 
1.1\%.

\section{Angular distribution in the $\pi\pi$ center-of-mass}

The pion angular distribution 
in the $\pi\pi$ center-of-mass with respect to the ISR photon direction in 
that frame, is model-independent. The $\cos\theta_\pi^*$ distribution behaves 
as $\sin^2\theta_\pi^*$ as a consequence of the P-wave between the two pions, 
but it is strongly distorted at $|\cos\theta_\pi^*|$ values near one due to the
$p>1$\GeVp~ cut on the tracks.

\begin{figure}[tb]
  \centering
\vspace{-0.5cm}
  \includegraphics[width=7.8cm]{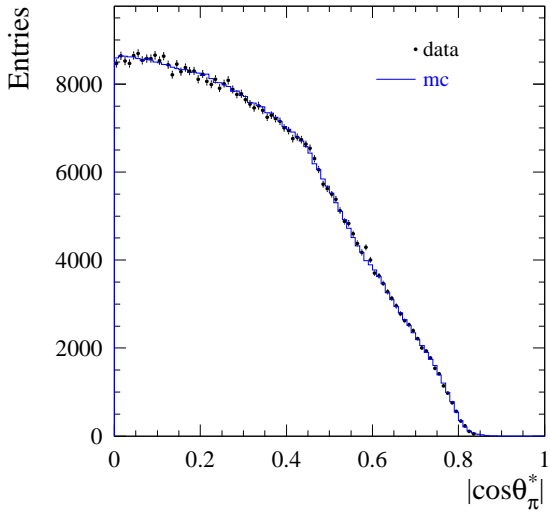}
\vspace{-0.5cm}
  \caption{\small 
The angular pion distribution in the $\pi\pi$ system with respect to the ISR
photon direction as function of $|\cos\theta_\pi^*|$ for background-subtracted
$\pi\pi\gamma(\gamma)$ data (points) in the $\rho$ central region
($0.5<m_{\pi\pi}<1$\GeVM). The blue histogram is the shape
obtained in the simulation, normalized to the data.}
  \label{cos-theta*-central}
\end{figure}

The $|\cos\theta_\pi^*|$ distributions for background-subtracted data 
and MC are compared in
Fig.~\ref{cos-theta*-central} for the $0.5-1$\GeVM~ mass range: they agree with
each other within the statistical uncertainties, as expected for a pure pion sample.

\section{The $\pi\pi$ cross section}

To correct for resolution and FSR effects, an unfolding of the background-subtracted $m_{\pi\pi}$ distribution, 
corrected for data/MC efficiency differences, is performed.
A mass-transfer matrix, created using simulation, provides the probability that an event generated in a $\sqrt{s'}$ 
interval $i$ is reconstructed in a $m_{\pi\pi}$ interval $j$. 
The matrix is corrected to account for the larger fraction of events with
bad $\chi^2$ values (and consequently poorer mass resolution) in data 
compared to MC because of the approximate simulation of additional ISR.  
The performance and robustness of the unfolding method~\cite{bogdan} 
have been assessed using test models. 

\begin{table} [tb] \centering 
\caption{ \label{syst-err} \small 
Relative systematic uncertainties (in $10^{-3}$) 
on the $e^+e^-\to\pi^+\pi^-(\gamma)$ 
cross section by $\sqrt{s'}$ intervals (in \GeVE) up to $1.2$\GeVE. 
The statistical part of the efficiency uncertainties is included in 
the total statistical uncertainty in each interval.}
\vspace{0.5cm}
{ \scriptsize 
\renewcommand{\tabcolsep}{0.15pc} 
\begin{tabular}{cccccc} 
\hline\hline 
Source of      & \multicolumn{5}{c}{$\sqrt{s'}$ (\GeVE)} \\
\cline{2-6}
 Uncertainty       &  0.3-0.4 & 0.4-0.5 & 0.5-0.6 & 0.6-0.9 & 0.9-1.2 \\ \hline
 trigger/ filter              & 5.3 & 2.7 & 1.9 & 1.0 & 0.5  \\ 
 tracking                     & 3.8 & 2.1 & 2.1 & 1.1 & 1.7  \\ 
 $\pi$-ID                     &10.1 & 2.5 & 6.2 & 2.4 & 4.2  \\
 background                   & 3.5 & 4.3 & 5.2 & 1.0 & 3.0  \\
 acceptance                   & 1.6 & 1.6 & 1.0 & 1.0 & 1.6  \\
 kinematic fit ($\chi^2$)     & 0.9 & 0.9 & 0.3 & 0.3 & 0.9  \\
 correlated $\mu\mu$ ID loss  & 3.0 & 2.0 & 3.0 & 1.3 & 2.0  \\
 $\pi\pi/\mu\mu$ non-cancel.  & 2.7 & 1.4 & 1.6 & 1.1 & 1.3  \\
 unfolding                    & 1.0 & 2.7 & 2.7 & 1.0 & 1.3  \\
 ISR luminosity ($\mu\mu$)    & 3.4 & 3.4 & 3.4 & 3.4 & 3.4  \\
\hline
 total uncertainty            &13.8 & 8.1 & 10.2 & 5.0 & 6.5  \\
\hline\hline
\end{tabular}
} 
\end{table}

The results for the $e^+e^-\to\pi^+\pi^-(\gamma)$ bare cross 
section including FSR, $\sigma^0_{\pi\pi(\gamma)}(\sqrt{s'})$, 
are given in Fig.~\ref{babar-log} (b).
Prominent features are the dominant $\rho$ resonance, the abrupt drop at 
$0.78$\GeVE~ due to $\rho-\omega$ interference, a clear dip at $1.6$\GeVE~ 
resulting from higher $\rho$ state interference,
and some additional structure near $2.2$\GeVE.
Systematic uncertainties are reported in Table~\ref{syst-err} for $0.3<\sqrt{s'}<1.2$\GeVE. 
Although larger outside this range, they do not exceed statistical errors over the full spectrum for the
chosen energy intervals.
In particular, a systematic uncertainty of only $0.5\%$ has been achieved in the central $\rho$ region. 

The BABAR data were compared to other experiments, exploiting a VDM fit of the pion form factor~\cite{bogdanSlides}. 
This fit describes well the BABAR data in the region of interest for the comparison. 
There is a relatively good agreement when comparing to the Novosibirsk data~\cite{cmd-2,snd} in the $\rho$ mass region, 
while a slope is observed when comparing to the KLOE '08 data~\cite{kloe08}. 
A flatter shape is observed when comparing to the recent KLOE~\cite{kloe10} data, obtained by the analysis of events 
with a detected, large angle ISR photon. 
A good agreement is observed when comparing to the Novosibirsk and KLOE data, in the low mass region, below $0.5$\GeVM. 
There is a good agreement between the BABAR data and the most recent $\tau$ data~(corrected for isospin-breaking effects) 
from Belle~\cite{belle}, while some systematic effects are observed when comparing to ALEPH~\cite{aleph05} and CLEO~\cite{cleo}. 

\section{The $\pi\pi$ contribution to $a_\mu$} 

The lowest-order contribution of the $\pi\pi(\gamma)$ intermediate state
to the muon magnetic anomaly is given by 
\begin{equation}
\label{eq:int_amu}
    a_\mu^{\pi\pi(\gamma),LO} \:=\: 
       \frac{1}{4\pi^3}\!\!
       \intl_{4m_\pi^2}^\infty\!\!\mathrm{d}s'\,K(s')\,\sigma^{0}_{\pi\pi(\gamma)}(s')~,
\end{equation}
where $K(s')$ is a known kernel~\cite{kernel}.
The integration uses the measured cross section and the
errors are computed using the full statistical and systematic covariance 
matrices. The systematic uncertainties for each source are taken to be fully 
correlated over the full mass region. 
The integrated result from threshold to $1.8$\GeVE~ is
\begin{equation}
    a_\mu^{\pi\pi(\gamma),LO} \:=\: (514.1 \pm 2.2 \pm 3.1)\times 10^{-10}~,
\end{equation}
where the errors are statistical and systematic.
This value is larger than that from a combination of previous 
$e^+e^-$ data~\cite{newtau} ($503.5\pm3.5$), but is 
in good agreement with the updated value from $\tau$ decay~\cite{newtau} 
($515.2\pm3.4$).
The deviation between the BNL measurement~\cite{bnl} and the theoretical prediction is reduced to $2.4$ standard deviations, 
when using the $\pi^+\pi^-$ data from BABAR only. 

\section{Other ISR measurements from BABAR} 

\begin{figure}[tb]
  \centering
  \includegraphics[width=7.8cm]{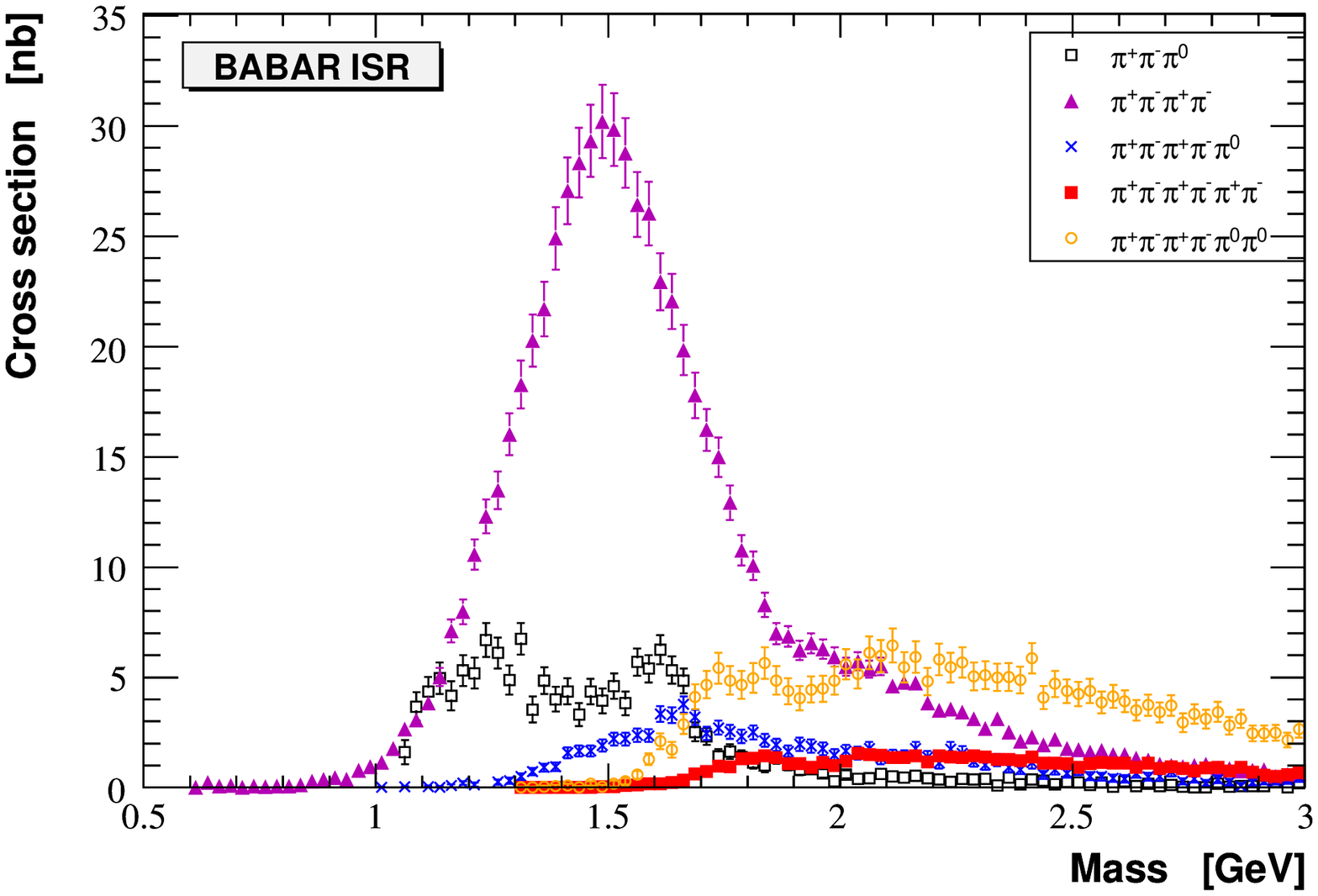} \\ 
  \includegraphics[width=7.8cm]{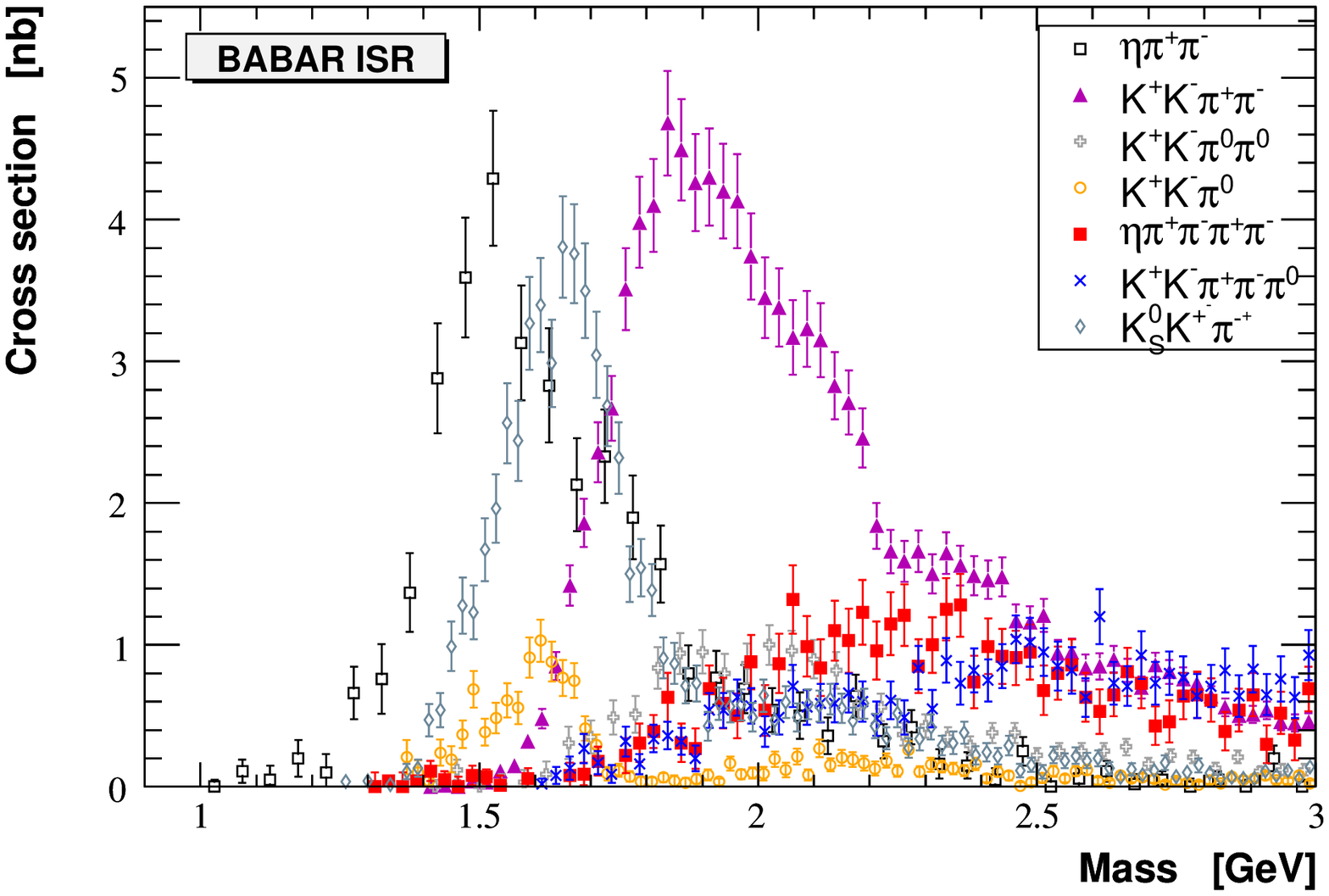} 
  \caption{\small 
Cross sections obtained by BABAR using the ISR method. The error bars include statistical and systematic uncertainties added 
in quadrature.} 
  \label{Fig:OtherISR_BABAR}
\end{figure}

Many other ISR measurements~\cite{babarMultiHad1,babarMultiHad2,babarMultiHad3,babarMultiHad4,babarMultiHad5}  
have been published by BABAR in the past (see Fig.~\ref{Fig:OtherISR_BABAR}). 
They have an improved precision comparing to previous measurements. 
Still more channels are under analysis: $K^+K^-$, $K\bar{K}\pi\pi$ including $K^0$'s, and $\pi^+\pi^-2\pi^0$. 
The last one is especially important in order to clarify the existing discrepancy between the $e^+e^-$ and $\tau$-based 
contributions to $a_\mu$ from this channel~\cite{Davier:2009zi}. 

\section{Conclusions and perspectives} 

BABAR has analyzed the $\pi^+\pi^-$ and $\mu^+\mu^-$ ISR processes in a consistent way, on the full mass range of interest ($0.3$ - $3$\GeVM). 
In addition, the absolute $\mu^+\mu^-$ cross section has been compared to the NLO QED prediction, the two being in agreement within $1.1\%$. 
The $e^+e^- \to \pi^+\pi^- (\gamma)$ cross section, obtained through the ratio of the $\pi^+\pi^-$ and $\mu^+\mu^-$ spectra 
is rather insensitive to the detailed description of radiation in MC. 
A strong point of the present analysis, comparing to previous ISR studies, comes from the fact that several uncertainties cancel 
in this ratio, allowing us to achieve our precision goal: 
the systematic uncertainty in the central $\rho$ region ($0.6-0.9$\GeVM) is only $0.5\%$. 

The contribution to $a_\mu$ from the BABAR $\pi^+\pi^-$ spectrum, in the range $0.28-1.8$\GeVE, 
is $(514.1 \pm 2.2 \pm 3.1)\times 10^{-10}$. 
This result has a precision of $0.7\%$, comparable to the combined previous measurements. 
The contribution from multi-hadronic channels will continue to be updated, with more results forthcoming from BABAR. 

In the comparison between the BABAR $\pi^+\pi^- (\gamma)$ cross section and the data from other experiments, 
there is a fair agreement with CMD2 and SND, while the agreement is poor when comparing with KLOE. 
The first priority should be to clarify the BABAR/KLOE discrepancy, the most important effect on $a_\mu$ being due to the difference 
on the $\rho$ peak. 
The origin of the slope in this comparison is also to be understood. 
The slope was very pronounced when comparing with the 2004 KLOE results, and it is reduced with the 2008 and 2010 data. 
The same slope is also observed in the comparison of the KLOE and $\tau$ data, while BABAR is in good agreement with the $\tau$ 
results. 
Further checks of the KLOE results are possible. 
Actually, as the method is based on MC simulation for ISR and additional ISR/FSR probabilities, 
an important cross-check should be provided by the comparison of the $\mu\mu$ spectrum with the QED prediction.

\end{document}